\documentclass{optica-article}

\journal{opticajournal} % for journals or Optica Open

\articletype{Research Article}

\usepackage{lineno}
\begin{document}

\title{Photonic Matrix Multiplier Makes a Direction-Finding Sensor}

\author{Kevin Zelaya\authormark{1} and Mohammad-Ali Miri\authormark{1,2*}}

\address{
\authormark{1} Queens College of the City University of New York, Department of Physics, Queens, New York 11367, USA\\
\authormark{2} Graduate Center of the City University of New York, Physics Program, New York, New York 10016, USA
}

\email{\authormark{*}mmiri@qc.cuny.edu} %% email address is required; see note below about the corresponding author designation

% use {asbstract*} to suppress the copyright line. Copyright information will be added in production

\begin{abstract*} 
We introduce a photonic integrated circuit solution for the direction-of-arrival estimation in the optical frequency band. The proposed circuit is built on discrete sampling of the phasefront of an incident optical beam and its analog processing in a photonic matrix-vector multiplier that maps the angle of arrival into the intensity profile at the output ports. We derive conditions for perfect direction-of-arrival sensing for a discrete set of incident angles and its continuous interpolation and discuss the angular resolution and field-of-view of the proposed device in terms of the number of input and output ports of the matrix multiplier. We show that while, in general, a non-unitary matrix operation is required for perfect direction finding, under certain conditions, it can be approximated with a unitary operation that simplifies the device complexity while coming at the cost of reducing the field of view. The proposed device will enable real-time direction-finding sensing through its ultra-compact design and minimal digital signal processing requirements.
%We introduce a lensless photonic integrated circuit solution for the direction-of-arrival estimation in the optical frequency band. The proposed circuit is built on discrete sampling of the phasefront of an incident optical beam and its analog processing in a photonic matrix-vector multiplier that maps the angle of arrival into the intensity profile at the output ports. We derive conditions for perfect direction-of-arrival sensing for a discrete set of incident angles and its continuous interpolation and discuss the angular resolution and field-of-view of the proposed device in terms of the number of input and output ports of the matrix multiplier. We show that while, in general, a non-unitary matrix operation is required for perfect direction finding, under certain conditions, it can be approximated with a unitary operation that simplifies the device complexity while coming at the cost of reducing the field-of-view. The proposed device will enable real-time direction-finding sensing through its ultra-compact design and minimal digital signal processing requirements.
\end{abstract*}

%%%%%%%%%%%%%%%%%%%%%%%%%%  body  %%%%%%%%%%%%%%%%%%%%%%%%%%
\section{Introduction}
Direction finding is a pivotal technology for identifying or determining the direction from which a received signal at radio or microwave frequencies originates. This practice is extensively used in a plethora of applications, including radar, wireless communications, and radio astronomy. The essential process involves the use of a radio receiver and antenna system to determine the spatial direction of an incoming radio signal. There are numerous methods applied for direction finding at RF and microwave frequencies~\cite{Yao2009microwave,perez2024general}, such as amplitude and phase comparison monopulse~\cite{sherman2011monopulse,ito2011antenna,feng2018mimo}, sequential lobing~\cite{lo1999theoretical,candan2015direction} (conical scanning and nutating), frequency interferometry~\cite{isleif2016experimental}, and time difference of arrival~\cite{vidal2006direction,Zou12,pertila2020time,zhang2023photonic}. In the optical frequency range, methods for direction finding often fall under the category of optical beamforming or beam steering, which is widely utilized in lidar (light detection and ranging) systems~\cite{poulton2017coherent,shang2017uniform,zhang2022large,sayyah2022fully,seyedinnavadeh2023determining}, free-space optical communications~\cite{poulton2017lens}, and various other applications~\cite{cheng2018recent}. In recent years, with the progress in developing densely packed photonic integrated circuits, numerous approaches have been undertaken to develop on-chip photonic lidar~\cite{chung202119,khachaturian2022discretization,khachaturian2022achieving} and remote sensing systems~\cite{White2020,serafino2020microwave}. 

In the past decade, there has been significant progress in developing programmable photonic integrated circuits capable of performing arbitrary discrete linear operations. This includes solutions based on meshes of Mach-Zehnder interferometers (MZI)~\cite{clements2016optimal, Shokraneh2020, rahbardar2023addressing}, multimode interference (MMI) couplers~\cite{pastor2021arbitrary}, and multiport waveguide arrays~\cite{tanomura_robust_2020,saygin_robust_2020, Skryabin2021,markowitz2023universal, Markowitz23Auto, zelaya2024goldilocks, markowitz2023learning}. The latter has been possible on platforms based on silicon solutions, such as SiO2 and SiN, where a high contrast refractive index is generated, rendering strongly confined and low-disperse guide modes propagating through waveguides~\cite{Siew2021}. In turn, tunable phase shifters based on thermo-optical~\cite{liu2022thermo} and phase-change materials~\cite{Rios2022ultra} (PCM) are implemented to program the device for the desired functionality. Given the scalability of these photonic circuits, programmable photonics has become an emerging and attractive technology that holds great promise for numerous applications in classical~\cite{bogaerts2020programmable} and quantum information processing~\cite{harris2017quantum,russell2017direct}. Despite the rapid development of programmable photonic integrated circuits, it remains to fully exploit novel functionalities for various use cases. Recent efforts have been put to deploy optical convolutional layers~\cite{meng2023compact,zelaya2024integrated}, neuromorphic optical computing~\cite{tait2017neuromorphic,shastri2021photonics}, and high-speed communication channels~\cite{shi2022silicon,shi2020scaling}.

In this work, we aim to focus on the direction-of-arrival finding and investigate the possibility of performing this fundamental task through a photonic matrix-vector multiplier circuit. We derive the required conditions and fundamental limitations so that a photonic circuit architecture for on-chip direction finding can be devised. The proposed architecture, shown schematically in Fig.~\ref{fig:fig1}, is based on a linear and equally spaced array of $M$ grating couplers that sample the phasefront of incoming plane waves. This discrete $M$-dimensional vector $\mathbf{x} \in \mathbb{C}^M$ is then steered into a photonic matrix-vector multiplier device that maps the phase slope of the incoming wave onto a localized intensity vector $\mathbf{y} \in \mathbb{C}^N$ at $N$ output ports. We consider and explore the general scenario in which the number of input ($M$) and output ports ($N$) is different. Thus, the photonic unit is, in general, non-unitary and is devised by imposing transformation rules on a set of incident sampling angles. This allows the transfer matrix of the device to be uniquely defined based on the design parameters, such as the number of grating couplers and their separation, as well as the detection aperture angle and the total number of output ports of the photonic processing unit. The detection functionalities of the device are proven to be continuously extended beyond the discrete sampling angles by introducing suitable tracking functions implemented in a post-processing stage at the output of the core photonic processor. The fundamental limits of each tracking function are discussed through pertinent examples. Lastly, an approximation is discussed so that the photonic operation reduces to a unitary one, rendering a compact device at the expense of compromises on the detection capabilities.
%%%%%%%%%%%%%%%%%%%%%%%%%%%%%%%%%%%%%%%%%%%%%%%%%%%%%%%%%%%%%%%%%%%%%%%%%%%
%--------------------------------------------------------------------------
\section{Theory}
\label{sec:theory}
\begin{figure*}
\centering
\includegraphics[width=0.8\textwidth]{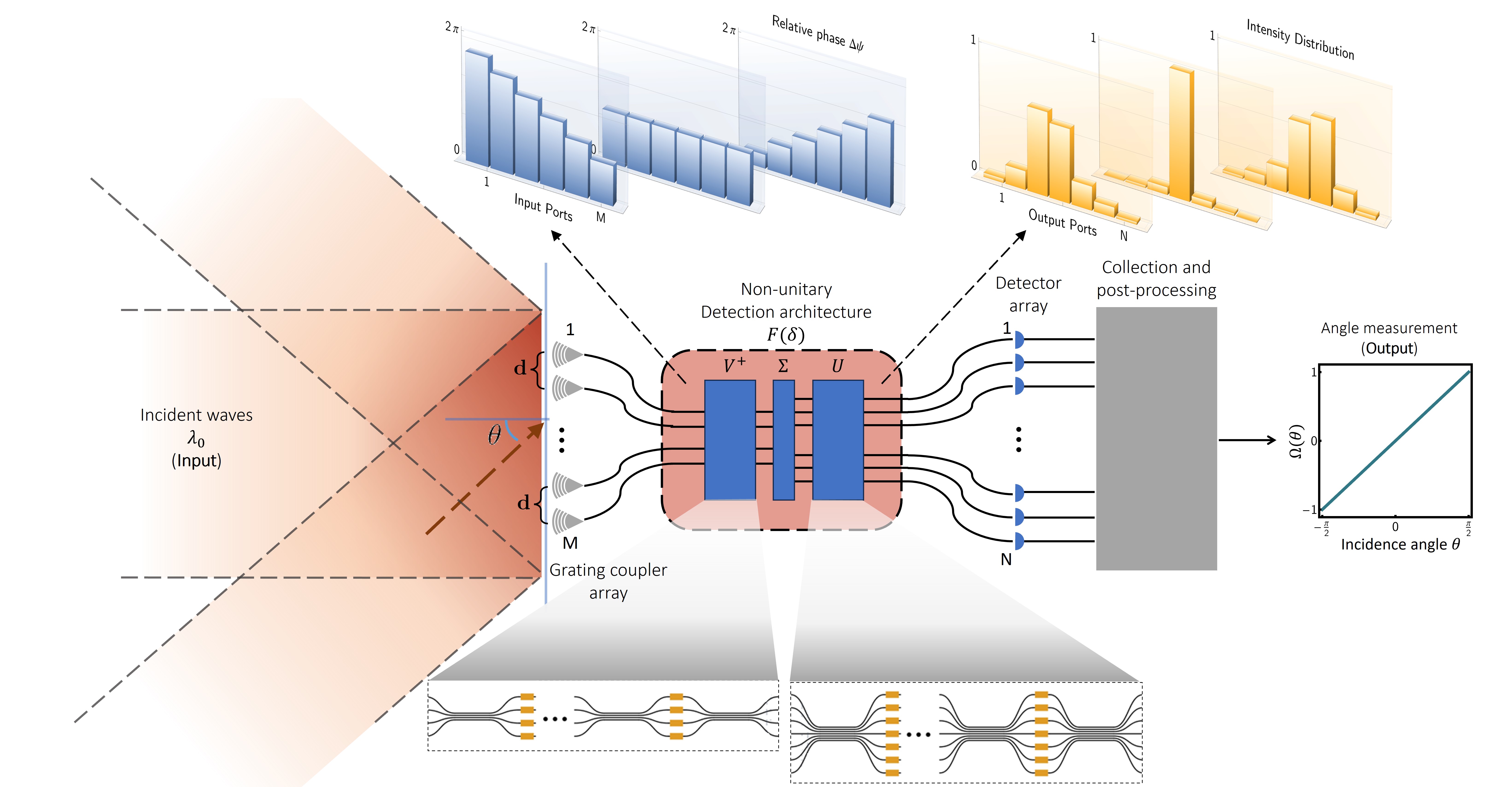}
\caption{The proposed device concept for direction-of-arrival sensing with a photonic analog matrix-vector multiplier as its core processor.}
\label{fig:fig1}
\end{figure*}
Let us consider an incoming plane wave with wavenumber $k_{0}$ traveling in the direction $\hat{\mathbf{r}}$. The wavefront is assumed to be incident onto an equally spaced linear array of $M$ grating couplers that take discrete samples of the plane wave. The distance between contiguous grating couplers is denoted by $d$, and the wavevector of the incoming plane wave makes an angle $\theta$ with respect to the linear array. These grating couplers collect the plane waves and steer them to a still unknown linear optical processor with transfer matrix $F$ that pre-processes the light gathered at the grating couplers and produces an $N$-port output $\mathbf{y}\in\mathbb{C}^{N}$, which is meant to be further processed to extract the features of the incident angle. The proposed architecture is sketched in Fig.~\ref{fig:fig1}. The optical processor is characterized by the linear operator $F\in\mathbb{C}^{N\times M}$, the explicit form of which is determined based on the input and output relations imposed on the system. 

We define the incident wave at a particular granting coupler as $e^{i \xi}/\sqrt{M}$, where the factor $1/\sqrt{M}$ is considered for normalization. Thus, straightforward calculations show that the contiguous grating coupler shall collect the wave $e^{i(\xi\pm k_{0}d\sin\theta)}/\sqrt{M}$. In this form, a vector $\mathbf{x}(\theta)$ containing the electric fields as measured at each of the $M$ spots on the grating coupler array, up to a global phase, is given by
\begin{equation}
\mathbf{x}(\theta)=\frac{1}{\sqrt{M}}\left(e^{-i\psi(\theta)}, e^{-2i\psi(\theta)}, \ldots, e^{-Mi\psi(\theta)} \right)^{T} ,
\label{x_th}
\end{equation}
where, $\psi(\theta):=k_{0}d\sin\theta$, with $\theta\in(-\pi/2,\pi/2)$, while $(\cdot)^{T}$ denotes the matrix transpose operation, and the factor $1/\sqrt{M}$ is for normalization.

The first stage of the device is then characterized by the set of linear operations $\mathbf{y}_{n}=F \mathbf{x}_{n}$, with $n=1,2,\ldots,N$ and $\mathbf{x}_{n}\in\mathbb{C}^{M}$ and $\mathbf{y}_{n}\in\mathbb{C}^{N}$ the respective input and output vectors of the transformation to be defined.

Here, we focus on a direction-finding device that, for a specific set of wavefronts with discrete incident angles $\{\theta_{n}\}_{n=1}^{N}$, produces a single impulse at the $n$-th output port of the photonic processor. Note that the number of sampling angles $\theta_{n}$ and the number of output ports is $N$, which is, in general, different from the number of grating couplers $M$. The previous requirements allow for the construction of a simple device that requires low computational power and performs a discrete detection operation, while we later discuss that the angle detection can be readily generalized to cover a continuous range. Since the grating couplers are linearly arranged, the maximum detection range (aperture angle) is $\pi$. For symmetry reasons, we choose the following discrete set of incident sampling angles:
\begin{equation}
\theta_{n}(\delta):=\left(-\frac{\pi}{2}+\delta\right)\left(\frac{N-2n+1}{N-1}\right) , \quad n\in\{1,\ldots,N\}.
\label{theta-n}
\end{equation} 
These angles are uniformly distributed in the interval $(-\pi/2+\delta,\pi/2-\delta)$, where $\delta$ controls the aperture angle for the detection scheme. Since we want to steer a specific incident wave with angle $\theta_{n}$ to the $n$-th output port, the direction-finding optical processor $F\equiv F(\delta)$ is devised through reverse engineering by imposing the set of input and output relations of the form
\begin{equation}
\hat{\mathbf{e}}_{n}=F(\delta)\mathbf{x}_{n}(\delta), \quad  n\in\{1,\ldots,N\}, 
\label{Fxn}
\end{equation}
where we have defined the set of \textit{discrete sampling vectors}
\begin{equation}
\mathbf{x}_{n}(\delta):=\mathbf{x}\left(\theta=\theta_{n}(\delta)\right), \quad  n\in\{1,\ldots,N\}, 
\label{sampling_xn}
\end{equation}
with $\hat{\mathbf{e}}_{n}=(0,\ldots,1,\ldots,0)^{T}$ the $n$-th vector of the canonical basis and $\mathbf{x}(\theta)$ is given in~\eqref{x_th}. To extract a simpler relation for $F(\theta)$, we exploit the fact that $\{\hat{\mathbf{e}}_{n}\}_{n=1}^{N}$ forms an orthonormal basis in $\mathbb{C}^{N}$. We thus multiply~\eqref{Fxn} by $\hat{\mathbf{e}}_{n}^{\dagger}$ to the right and sum over $n$ to render the relation
\begin{equation}
\mathbb{I}_{N}=F(\delta)G(\delta), \quad G(\delta):=\sum_{n=1}^{N}\mathbf{x}_{n}\hat{\mathbf{e}}_{n}^{\dagger}\equiv 
\left( \mathbf{x}_{1}(\delta), \ldots, \mathbf{x}_{N}(\delta) \right) .
\label{FG}
\end{equation}
where $\mathbb{I}_{N}$ is the identity matrix in $\mathbb{C}^{N\times N}$ and $G(\delta)\in\mathbb{C}^{M\times N}$. Nevertheless, the exact form and solvability of $F(\delta)$ is constrained to the exact relation between the number of output ports ($N$) and grating couplers ($M$).

\begin{figure}
    \centering
    \includegraphics[width=0.45\textwidth]{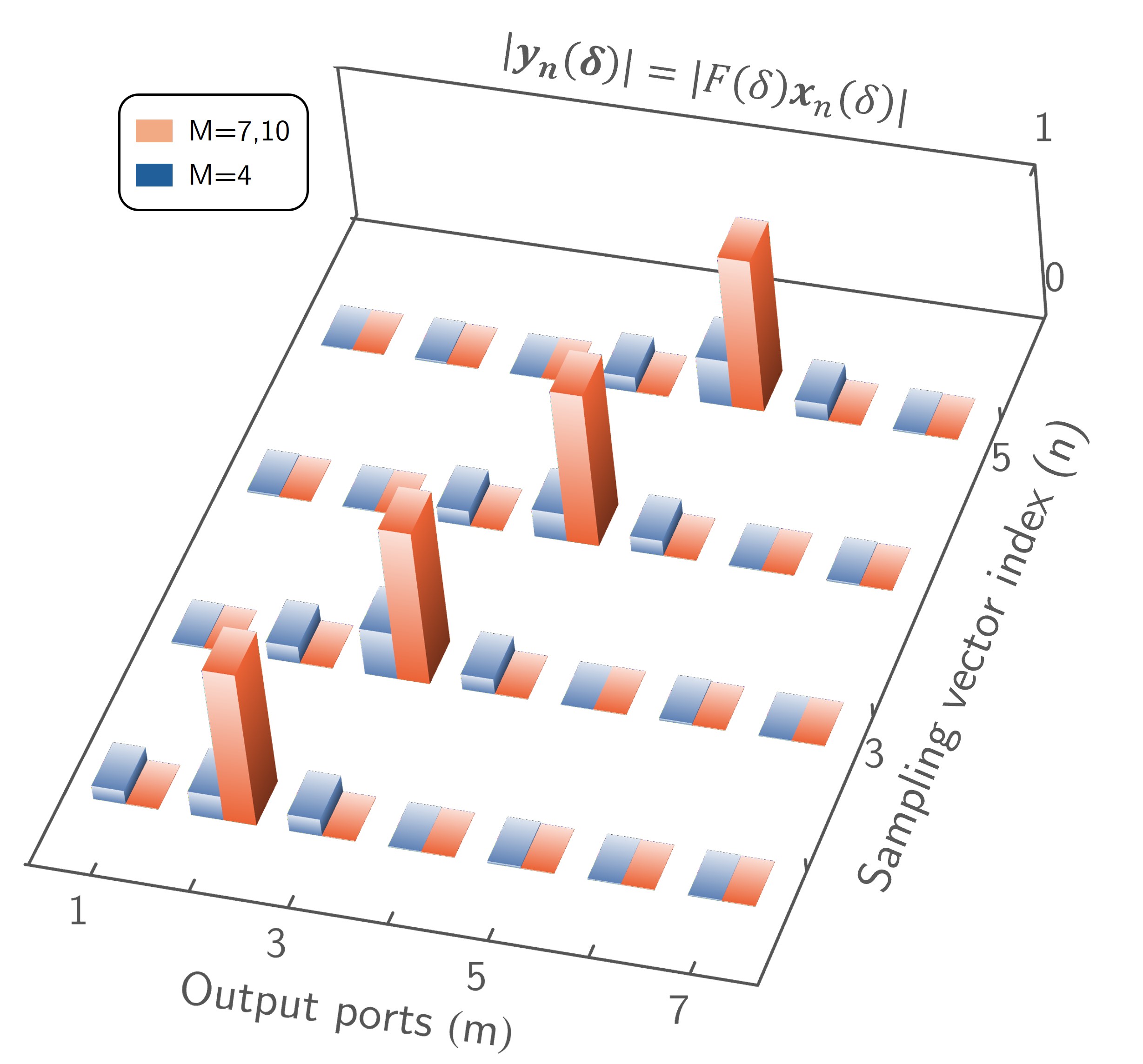}
    \caption{Output intensity distribution $\vert y_{n;m}\vert^{2}$, with $\mathbf{y}_{n}:=(y_{n;1},\ldots,y_{n;N})\equiv F(\delta)\mathbf{x}_{n}(\delta)$, for $N=7$ with $M=7,10$ (red) and $M=4$ (blue), along with $\kappa_{0}d=1$.}
    \label{fig:out_NM}
\end{figure}

For $M=N$, $G(\delta)$ reduces to a non-unitary square matrix. The non-unitarity becomes evident as the column vectors $\mathbf{x}_{n}$ are not orthonormal. However, the invertibility is ensured as $G(\delta)$ takes the form of a \textit{non-uniform discrete Fourier transform}~\cite{Bagchi01} (NUDFT), alternatively known as Vondermonde matrices~\cite{Orucc07,Moon00}. From this, the existence and uniqueness of the inverse of $G(\delta)$ is secured since the determinant of Vondermonde matrices~\cite{Orucc07} is always non-null. Thus, $F(\delta)$ always exists for $N=M$, and our construction ensures that the output will exactly reproduce an excitation in one single output port for plane waves incident at the exact angle $\theta_{n}$.

In turn, for $M\neq N$, the transforms $F$ and $G$ are defined by rectangular matrices whose inverses either do not exist or are not uniquely defined. This issue is overcome by considering the pseudo-inverse, also known as the Moore-Penrose inverse~\cite{penrose1955generalized}, denoted by $G^{-}$. The latter fulfills the property $G G^{-} G=G$, which holds for square nonsingular matrices, as well as for left-inverse and right-inverse matrices for rectangular matrices. Note that Eq.~\eqref{FG} defines a system of $N^{2}$ equations and $NM$ unknown variables $F_{p,q}$, which leads to an over-parameterized and an under-parameterized problem form $M<N$ and $M>N$, respectively. The former has more than one solution, whereas the latter does not have a solution. Here, the pseudo-inverse $G^{-}$ allows for a solution~\eqref{FG}, for its existence and uniqueness are ensured and can be determined from the singular value decomposition (SVD) of the matrix in question. That is, $G=U\Sigma V^{\dagger}$, with $U$ and $V$ unitary matrices and $\Sigma$ a diagonal matrix containing the singular values of $G$. For $M<N$, $G^{-}$ provides a particular exact solution. For $M>N$, $G^{-}$ yields the best approximate solution that minimizes the involved least square problem~\cite{golub2013matrix}. Therefore, in all cases, the liner operator characterizing the device is
\begin{equation}
\label{F_device}
F(\delta)=G^{-}(\delta), \quad G^{-}(\delta)=V\Sigma^{-}U^{\dagger},
\end{equation}
where $U$ and $V$ are computed from the SVD of $G(\delta)$, and $\Sigma^{-}$ is the pseudo-inverse of $\Sigma$. To illustrate the two cases posed above, we illustrate in Fig.~\ref{fig:out_NM} the out output mode $\mathbf{y}_{n}=F\mathbf{x}_{n}$ for $N=7$ and $M=4,10$, and grating couplers separation in the sub-wavelength domain $d_{0}=\lambda_{0}/2\pi$. In such a case, the output renders the desired operation $\mathbf{y}_{n}=\hat{\mathbf{e}}_{n}$ for $M>N$, as expected. In turn, one might notice a distribution for $M<N$ whose peak is located at the $n$-th port. This solution is the best approximation for the under-parameterized set of equations, which still captures part of the direction-finding operation. Lastly, for $M=N$, the pseudo-inverse reduces to the conventional inverse discussed above.

%
%--------------------------------------------------------------------------
%%%%%%%%%%%%%%%%%%%%%%%%%%%%%%%%%%%%%%%%%%%%%%%%%%%%%%%%%%%%%%%%%%%%%%%%%%%

%%%%%%%%%%%%%%%%%%%%%%%%%%%%%%%%%%%%%%%%%%%%%%%%%%%%%%%%%%%%%%%%%%%%%%%%%%%
%--------------------------------------------------------------------------
\section{Angle-tracking operations}
\begin{figure*}
\centering
\includegraphics[width=0.8\textwidth]{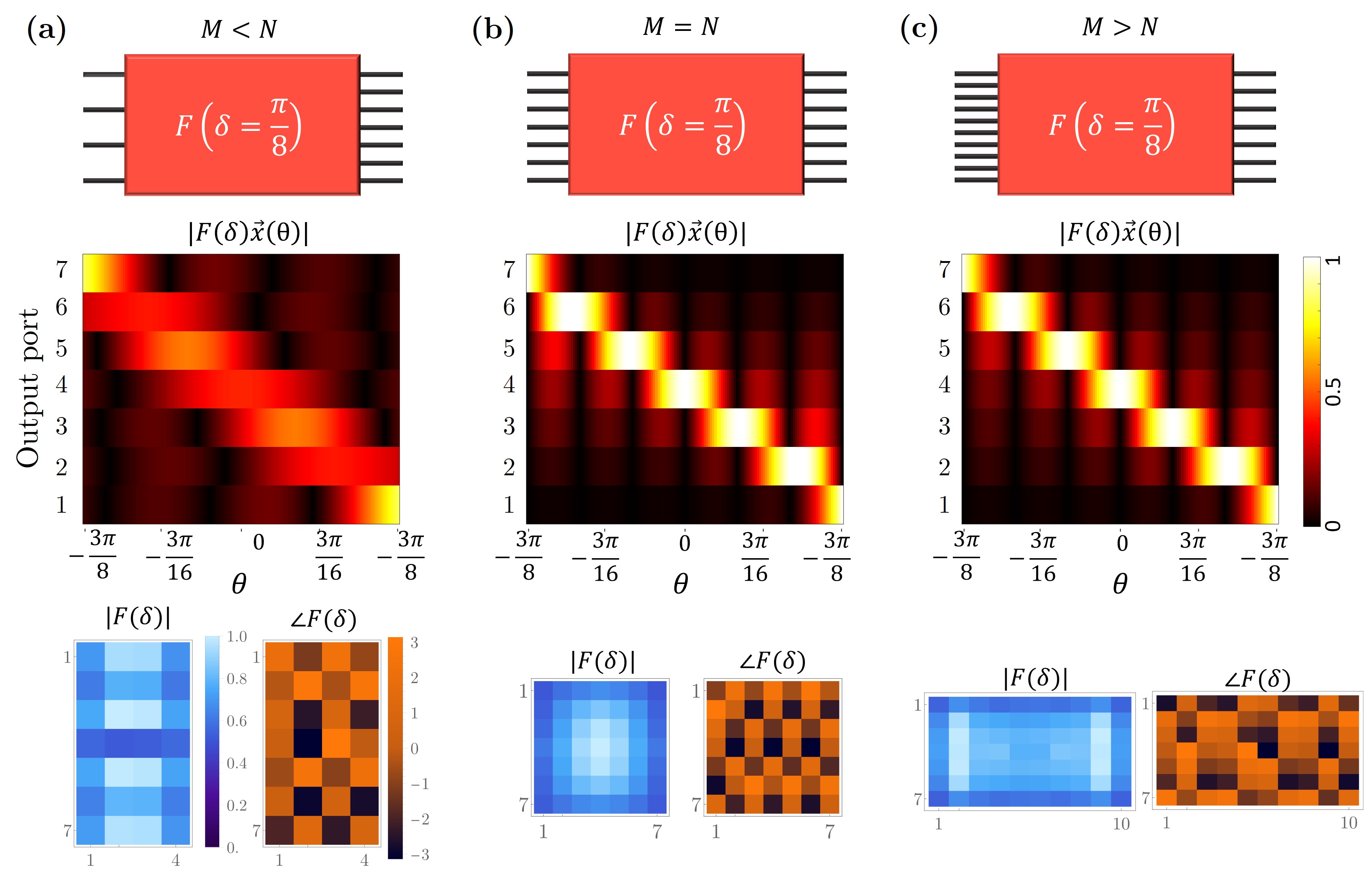}
\caption{Output intensity distribution $\vert F(\delta)\mathbf{x}(\theta)\vert$ for $N=7$ and $M=4$ (a), $M=7$, and $M=10$. The parameters are $\delta=\pi/8$, $\ell=1/2\pi$ ($\kappa_{0}d=1$), and waves with incident angles $\theta\in(-3\pi/8,3\pi/8)$.}
\label{fig:distribution_1}
\end{figure*}
The operator $F(\delta)$ has been constructed so that the set of discrete incident angles $\mathbf{x}_{n}$ can be exactly detected at the output. However, incident waves at the grating couplers arrive at continuous angles. Thus, to keep a better track of the incident wave angle, a post-processing operation on the signal generated by $F(\delta)$ is required so that the detection capabilities of the device extend beyond the $N$ incident angles $\mathbf{x}_{n}$. To this end, it is convenient to inspect $F(\delta)\mathbf{x}(\theta)$ for a continuum interval $\theta\in(-\pi/2+\delta,\pi/2-\delta)$. Such an operation is portrayed in Fig.~\ref{fig:distribution_1} for $N=7$ and a grating coupler separation $d=\lambda_{0}\ell=1$, with $\ell=1/2\pi$, and the three possible cases $M=4,7,10$. From the latter, one may note that the intensity distribution has a maximum whose location moves from the first output port ($n=1$) to the last one ($n=N=7$). Additionally, for a fixed $\theta$, the output distribution has small tails around such the maximum, resembling the shape of a probability distribution. The latter is not necessarily normalized as $F(\delta)$ defines a non-unitary transform, even for $M=N$. Likewise, a similar behavior is found for $M>N$, where the pseudo-inverse $G^{-}$ is exact. Although the distribution spreads more for $M<N$, where the pseudo-inverse is only approximated, the overall dynamic is similar to the other cases. 

The latter suggests that some aperture detection angles can be detected by either analyzing the intensity distribution across the output ports or pre-processing it before its detection. One way to achieve continuous tracking of the incident wave angle $\theta$ is by defining the position vector $\mathbf{q}:=(1,\ldots,N)^{T}$ and computing the ``average position'' defined through the matrix-vector multiplication
\begin{equation}
\Omega(\theta;\ell):=\vert\mathbf{q}^{\dagger}F(\delta;\ell) \mathbf{x}(\theta)\vert , \quad \mathbf{q}=(1,2,\ldots,N)^{T} , \quad d=\ell\lambda_{0},
\label{angle-cont}
\end{equation}
with $\mathbf{q}^{\dagger}$ the complex transpose (adjoint) of $\mathbf{q}$. In the latter, we have rescaled the grating coupler spacing in terms of the incident wavelength $\lambda_{0}=2\pi/\kappa_{0}$ through the relation $d=\ell\lambda_{0}$, where $\ell$ defines the spacing factor. From now on, we refer to $\Omega(\theta;\ell)$ as the \textit{continuous tracking measure}. Remark that incident angles in the set $\theta\in\{\theta_{n}\}_{n=1}^{N}$ yield to  $\Omega(\theta_{n};\ell)=n$; i.e., the continuous tracking measure leaves invariant the original detection angles scheme of the sampling angles $\mathbf{x}_{n}$. To make a reliable prediction of the incident angle out of the continuous tracking function, the $\Omega(\theta;\ell)$ shall define a one-to-one function so that no ambiguity exists in the output measurement with respect to the incident angle $\theta\in(-\pi/2+\delta,\pi/2-\delta)$ for a fixed grating coupler separation $d=\lambda_{0}\ell$. Otherwise, the angle detection would be ambiguous and thus ill-defined.

Alternatively, a second measure can be established to track the incident wave angle. As pointed out above, the maximum of the intensity distribution $\vert F(\delta;\ell)\mathbf{x}(\theta)\vert$ moves in discrete steps along the incident angle $\theta$. This suggests an alternative operation of the form 
\begin{equation}
\Phi(\theta;\ell)= max(\vert F(\delta;\ell)\mathbf{x}(\theta) \vert) ,
\label{angle-disc}
\end{equation}
where $max(\vert\mathbf{z}\vert)$ computes the position $n\in\{1,\ldots,N\}$ in which the maximum element $\vert z_{n}\vert$ of $\vert\mathbf{z}\vert$ is located. Clearly, this measure defines a discrete mapping $\Phi(\theta;\ell):\mathbb{C}^{N}\mapsto\mathbb{N}$, which also preserves the output for the sampling discrete angles $\{\theta_{n}\}_{n}^{N}$; that is, $\Phi(\theta_{n};\ell)=n$, just like the continuous tracking measure does. Henceforth, we refer to $\Phi(\theta;\ell)$ as the \textit{discrete tracking measure}.

\begin{figure*}
    \centering
    \includegraphics[width=0.9\textwidth]{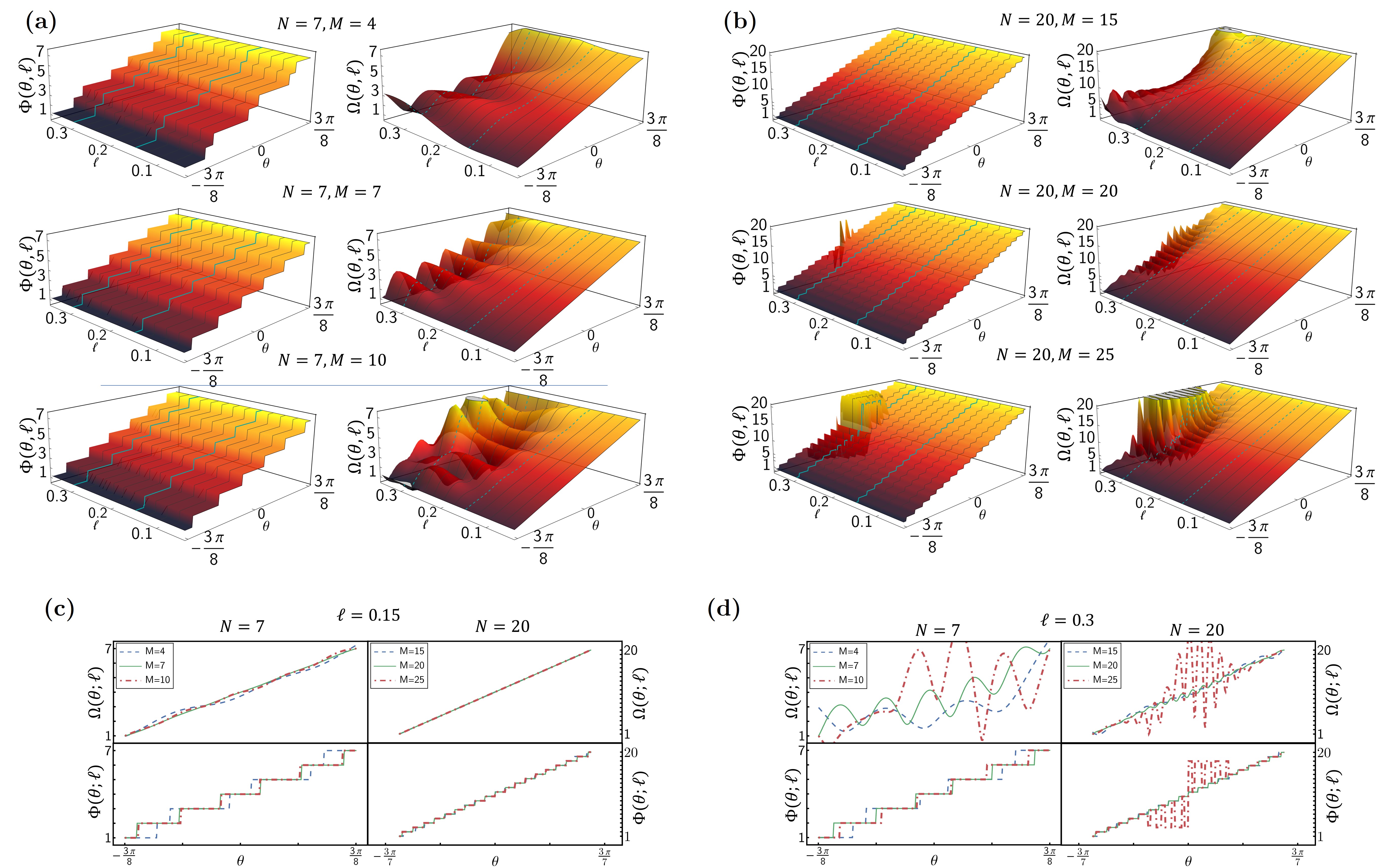}
    \label{fig:omega-angles-d}
    \caption{Continuous ($\Omega(\theta;\ell)$) and discrete ($\Phi(\theta;\ell)$) tracking functions for incident waves $\mathbf{x}(\theta)$ as a function of the spacing factor $\ell$ ($d=\lambda_{0}\ell$) and incidence angles $\theta$. 3D plots depict both tracking functions for $\{N=7,M=4,M=7,M=10\}$ (a) and $\{N=20,M=15,M=20,M=25\}$ (b). Corresponding projections at $\ell=0.15$ (c) and $\ell=0.3$ (d).}
    \label{fig:omega-phi-angles}
\end{figure*}

Although the device operator $F(\delta)$ can be precisely constructed for any combination of the parameter set $\{\ell,M,N\}$, the continuous and discrete tracking functions might not be suitable for all arbitrary values of such parameters. Some compromises might arise when dealing with different spacing factors $\ell$. To this end, Fig.~\ref{fig:omega-phi-angles}(a)-(b) illustrates the behaviour of the $\Omega(\theta;\ell)$ and $\Phi(\theta;\ell)$ as a function of both the incident wave angle $\theta$ and the spacing factor $\ell$. Here, we focus on $\ell<1$, which defines grating couplers spaced in the sub-wavelength regime and output ports $N=5$ and $N=20$. For $N=7$, the discrete measure works as required for $\ell\in(0.05,0.31)$ regardless of the values of $M$. Despite the latter, the resolution of the measure is relatively poor, as only seven angles can be reported. In turn, the continuous measure can be used indeed, but only for limited cases of the spacing factor $\ell\lessapprox 0.15$. For larger values of $\ell$, $\Omega(\theta;\ell)$ ceases to be a one-to-one function and is thus discarded as a candidate for tracking function. This is portrayed in Fig.~\ref{fig:omega-phi-angles}(c). 

The situation improves by increasing the number of output ports to $N=20$, where $\Omega(\theta;\ell)$ becomes a well-posed tracking function for $\ell\lessapprox 0.28$ when $M=20$ and $M=15$, and similarly for $\ell\lessapprox 0.25$ when $M=25$. For clarity, the reliability is further illustrated in Fig.~\ref{fig:omega-phi-angles} for $\ell=0.15$ and $\ell=0.3$. For $\ell=0.3$ and $N=7$, $\Omega(\theta;\ell)$ shall be ruled out for any $M$ and be replaced by $\Phi(\theta;\ell)$ instead. Likewise, for $\ell=0.3$ and $N=20$, both $\Omega(\theta;\ell)$ and $\Phi(\theta;\ell)$ are suitable except for $M=25$. This provides some guidelines for the design of the final device based on the accuracy and compactness requirements.

%
%--------------------------------------------------------------------------
%%%%%%%%%%%%%%%%%%%%%%%%%%%%%%%%%%%%%%%%%%%%%%%%%%%%%%%%%%%%%%%%%%%%%%%%%%%

%%%%%%%%%%%%%%%%%%%%%%%%%%%%%%%%%%%%%%%%%%%%%%%%%%%%%%%%%%%%%%%%%%%%%%%%%%%
%--------------------------------------------------------------------------
\section{Device concept and design}
\label{sec:device}
The all-optical implementation for the direction finder architecture described by the non-unitary matrix $F(\delta)$ can be determined, in general, by using singular value decomposition (SVD), which in each case is straightforwardly obtained by numerical means. This is done for every configuration, leading to the factorization
\begin{equation}
F=U \Sigma V^{\dagger},
\label{SVD}
\end{equation}
with $U\in\mathbb{C}^{N\times N}$ and $V\in\mathbb{C}^{MM}$ unitary matrices. In turn, $\Sigma\in\mathbb{C}^{M\times N}$ is a semi-positive definite rectangular diagonal matrix. 

The factorization~\eqref{SVD} makes it possible to design an equivalent photonic circuit that captures all the properties of the direction-finding element. On the one hand, the two unitary elements can be constructed either through meshes of MZI~\cite{Reck94,Clements16,miller2013self} or arrays of coupled waveguides~\cite{tanomura_robust_2020,Markowitz23,Markowitz23Auto}. In turn, the middle section of the architecture $\Sigma$ involves a semi-positive rectangular diagonal matrix, which can be deployed using a $N$-parametric layer of MZI interferometers. Although each MZI requires two-phase elements to steer both amplitude and phase, one MZI is required in the current approach since the architecture involves only amplitude modulation. Alternatively, amplitude modulation is achievable by utilizing phase-change materials (PCM), which allows for a more compact solution as compared to the MZI array~\cite{Rios2022ultra,youngblood2023integrated,zhang2019broadband}. If $\ell$ is a fixed quantity due to the non-tunability of the grating coupler separation, the transmission matrix $F(\delta)$ is also fixed. This permits designing the corresponding photonic circuit so that no active elements are needed. This holds if we use MZI as the modulation layer, whereas for PCMs, there is still a power consumption requirement to operate them. Here, the unitaries $U$ and $V^{\dagger}$ are implemented using interlaced layers of phase shifters $P^{(n)}$ and waveguide arrays $G$ through the universal factorization~\cite{Markowitz23Auto}
\begin{equation}
\begin{aligned}
&\mathcal{U}=GP^{(K+1)}G\ldots G P^{(1)} G , 
\\& G=e^{-iH L}, \quad P^{(n)}_{p,q}=e^{i \phi_{p}^{(n)}}\delta_{p,q},
\end{aligned}
\label{univ_U}
\end{equation}
where $L$ is the coupling length of the waveguide array, $\phi^{(n)}_{p}\in(0,2\pi)$ is the p-th phase element in the n-th layer, $p,q\in\{1,\ldots,K\}$ and $K\in\mathbb{Z}^{+}$. Here, $K$ is the dimension of the corresponding target unitary matrix $\mathcal{U}$, which can be either N  and M for $U$ and $V^{\dagger}$, respectively. The tridiagonal matrix $H\in\mathbb{R}^{K\times K}$ corresponds to the JX lattice Hamiltonian~\cite{Wei16}, with components $H_{p,q}=\kappa_{p+1}\delta_{p+1,q}+\kappa_{p}\delta_{p-1,q}$ for $p,q\in\{1,\ldots,K\}$. 

\begin{figure*}
    \centering
    \includegraphics[width=0.8\textwidth]{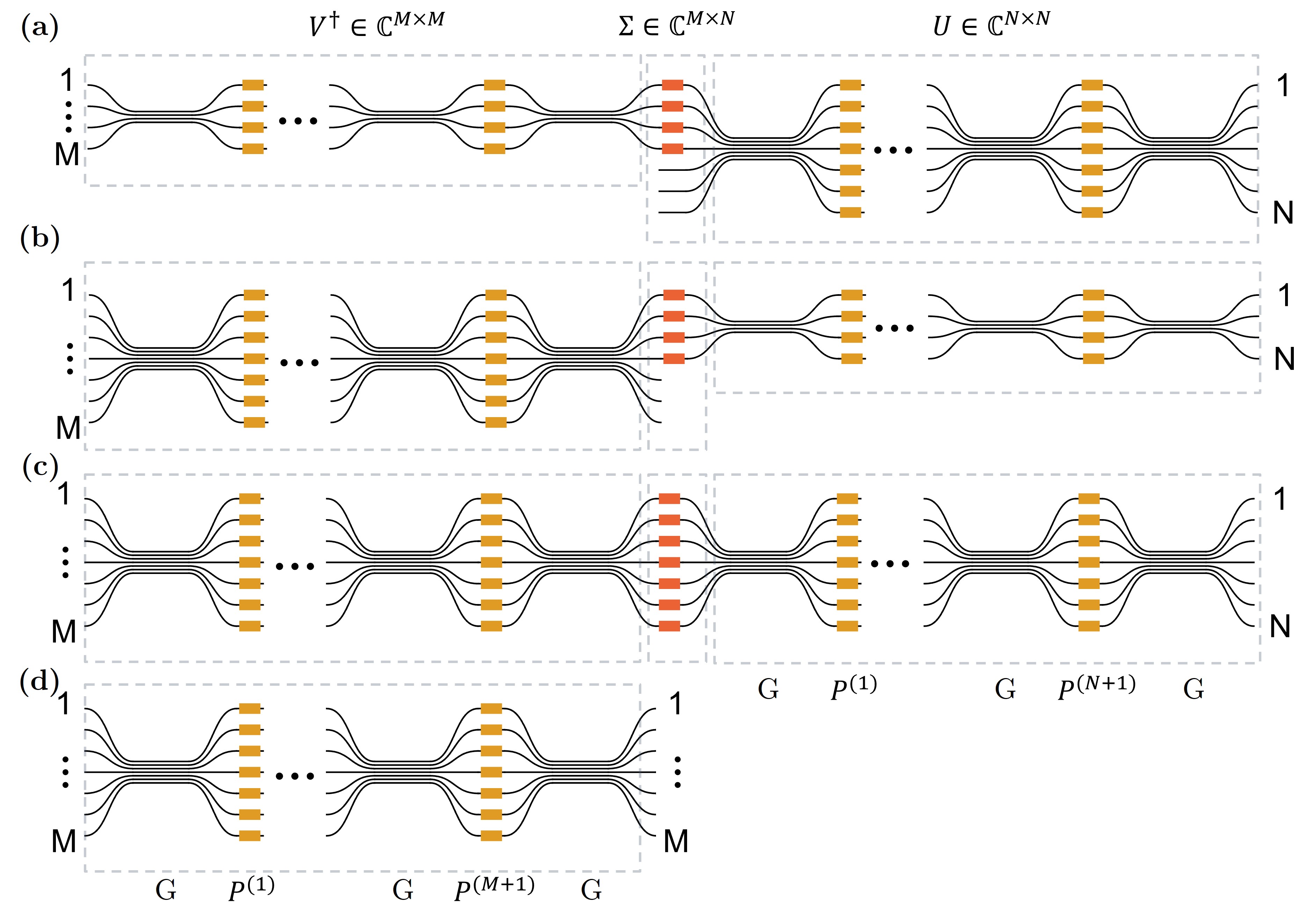}
    \caption{Skecth of the architecture characterizing the linear transform $F(\delta)$ for $M<N$ (a), $M>N$ (b), and $M=N$ (c). In all cases, the yellow and red blocks denote the phase and amplitude modulators, respectively. The corresponding architecture for the unitary limit is depicted in panel (d).}
    \label{fig:device}
\end{figure*}

Sketches of the photonic circuits related to the architecture $F(\delta)$ are depicted in Fig.~\ref{fig:device}. The yellow blocks denote the phase elements, and the red blocks are the amplitude modulation elements. Particularly, Figs.~\ref{fig:device}(a)-(b) show the cases $M<N$ and $M>N$, respectively, where clearly not all ports of the unitary matrices $V^{\dagger}$ and $U$ are connected. This suggests that $F(\delta)$ can be rendered using lesser elements in one of the unitaries. Indeed, the universality of Eq.~\eqref{univ_U} was numerically shown for $K(K+1)$ phase elements~\cite{Markowitz23,Markowitz23Auto}. Here, for $N\neq M$ and from the SVD, one may note that fewer phase elements are required. To illustrate this, recall that $\Sigma$ is a $M\times N$ rectangular diagonal matrix with $(N-M)$ zero columns and $(M-N)$ zero rows for $M<N$ and $M>N$, respectively. This implies that only the first $M$ rows of $V^{\dagger}$ ($M<N$) and $U$ ($N>M$) are meaningful for the reconstruction of $F(\delta)$, and the resulting unitary architectures shall require less optical elements.

%
%---------------------------------------------------------------------
\subsection{Unitary approximation}
Further simplifications are possible for $N=M$ since unitary matrices can be recovered from special considerations. This is a rather ideal scenario, as the number of elements of the proposed architecture further simplifies to $N^2$. Furthermore, all eigenvalues of $F(\delta)$ are unimodular in the unitary limit, allowing us to bypass the use of the amplitude modulators and utilize an architecture that requires only one unitary processor, i.e.,
\begin{equation}
   F = U \Sigma V^{\dagger} \approx U V^{\dagger}=\widetilde{U}\in U(N).
\end{equation}
Such a limit can be addressed by considering the cases in which the NUDFT reduces to a DFT, which is indeed unitary. From~\eqref{FG} and~\eqref{F_device}, for $M=N$, one may notice that the non-unitarity of $F(\delta)$ arises from the non-uniform sampling of the function $\psi(\theta_{q})$ across $N$ equally spaced points. We shall look for cases where $\psi(\theta_{q})$ can be expanded, within some degree of accuracy, as a linear function of $\theta_{q}$, which is in turn a linear function of $q$. 

Let us consider the particular case where the aperture angle $(-\pi/2+\delta,\pi/2-\delta)$ is small enough so that $\psi(\theta_{n}(\delta))=\kappa_{0}d\sin\theta_{n}(\delta)\approx\kappa_{0}d\theta_{n}(\delta)$, for all $n\in\{1,\ldots,N\}$. Since $\theta_{n}$ in~\eqref{theta-n} are symmetrically distributed, one has $\vert\theta_{1}\vert=\vert \theta_{N}\vert>\vert\theta_{n}\vert$, for $n\in\{2,\ldots,N-1\}$. This reduces the condition for the unitary approximation to $\vert \theta_{1}\vert\ll 1$, which can be further simplified by introducing the half-aperture angle\footnote{Note that $2\epsilon$ defines the total detection aperture angle.} $\epsilon=\pi/2-\delta\geq 0$, yielding to the relation $\epsilon\ll 1$. Although this approximation does not restrict the total number of ports $N$, it compromises the detection range capabilities of the current architecture. Under such circumstances, the matrix elements of the linear transform~\eqref{F_device} are reduced to $F_{p,q}(\delta;\ell)\approx \frac{1}{\sqrt{N}}\exp\left( \frac{4\pi i \ell \epsilon}{N-1}p q-2\pi i \ell \epsilon \frac{N+1}{N-1}p \right)$. The straightforward calculations show that the latter defines a unitary matrix if the second constraint $\ell=\frac{1}{2\epsilon}\left(1-\frac{1}{N}\right)$ is imposed. This establishes a relation between the spacing factor $\ell$ and the aperture angle $2\epsilon$, where $\epsilon$ shall be small enough and non-null to avoid a null aperture angle. These conditions render the approximated unitary matrix $\widetilde{F}^{(u)}$ with matrix components
\begin{equation}
\widetilde{F}^{(u)}_{p,q}:=\frac{1}{\sqrt{N}}\exp\left( \frac{2\pi i q }{N} \left( p -\frac{N+1}{2} \right) \right).
\label{unitary-1}
\end{equation}
The photonic circuit related to this unitary device is then implemented using only one unitary block~\eqref{univ_U}, as depicted in Fig.~\ref{fig:device}(d).

To test the accuracy of the unitary approximation, one may first inspect the deviations in percentage error between $\sin(\theta_{1})\equiv \sin(\epsilon)$ and $\epsilon$, which leads to the errors 2.61$\%$ and 11,07$\%$ for the half-aperture angles $\epsilon=\pi/8$ and $\epsilon=\pi/4$, respectively. The latter provides some preliminary information on the performance of the unitary approximation as a function of $\epsilon$. This is indeed illustrated by computing the distance between the approximated unitary matrix~\eqref{unitary-1} and the exact expression~\eqref{FG} using the Frobenius norm for the allowed values of $\ell$ and $\epsilon$. For completeness, we also compute the distance between the singular values $\Sigma$ and the identity matrix, for in the unitary approximation, the former one shall be close to the identity, reducing the SVD into the product of two unitary matrices $F=U\Sigma V^{\dagger}\approx U V^{\dagger}$. These two distances are respectively given by
\begin{equation}
d_{F}=\frac{1}{N^{2}}\Vert \mathbb{I}-FF^{\dagger}\Vert^{2}_{F} , \quad d_{\Sigma}=\frac{1}{N^{2}}\Vert \mathbb{I}-\Sigma^{2}\Vert^{2}_{F} ,
\end{equation}
where $\Vert \cdot\Vert_{F}$ stands for the Frobenius norm. 

\begin{figure*}
    \centering
    \includegraphics[width=0.8\textwidth]{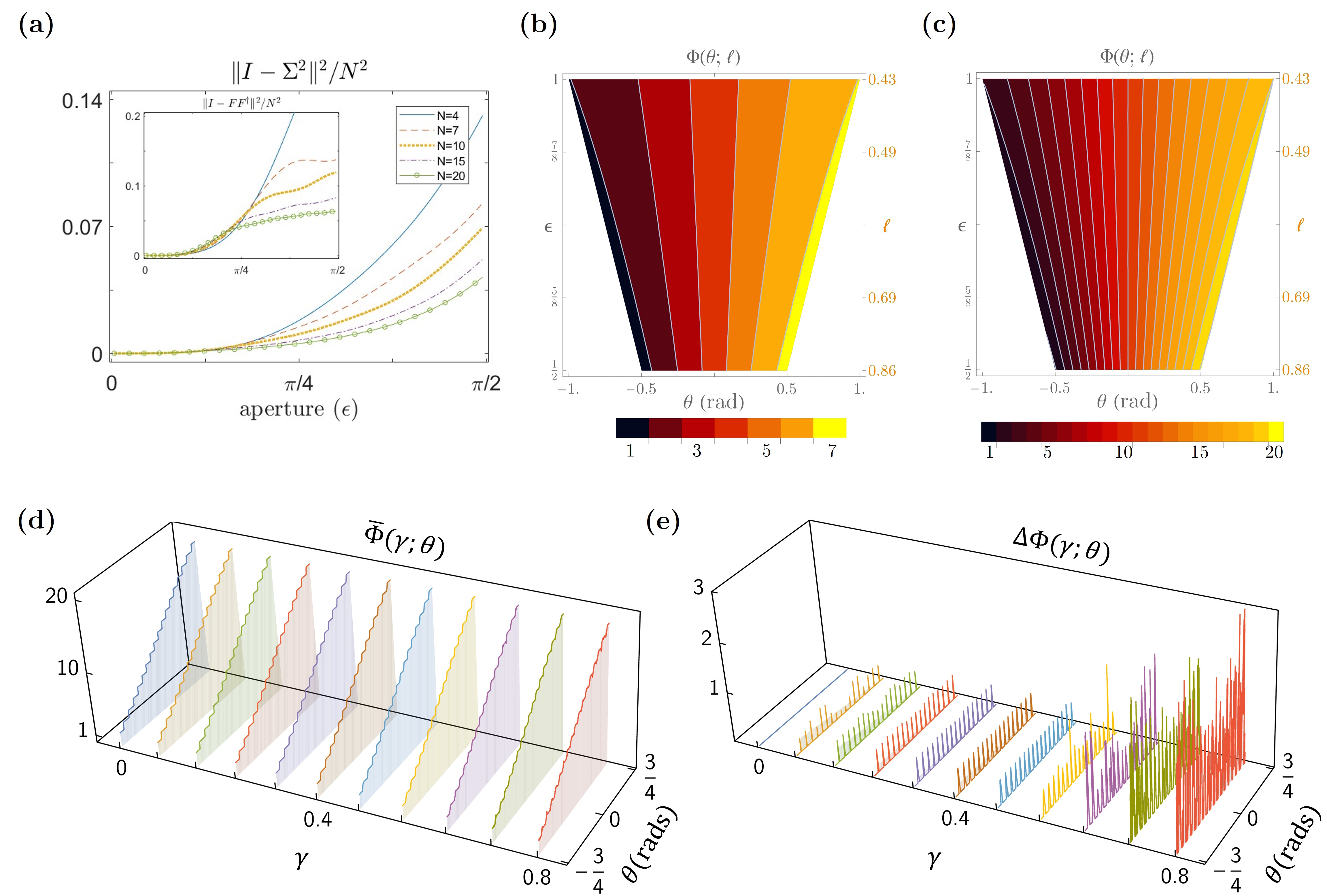}
    \caption{(a) Distance functions $d_{\Sigma}(\epsilon,N)$ and $d_{F}(\epsilon,N)$ (inset) as a function of the aperture angle $\epsilon$ for several total number of ports $N$. (b)-(c) Density plot of the discrete tracking function $\Phi(\theta;\ell)$ for $N=7$ (b) and $N=20$ (c) in the unitary approximation as a function of $\theta\in(-\epsilon,\epsilon)$ and $\epsilon\in(0.5,1)$ (or equivalently $\ell$).}
    \label{fig:case_unitary}
\end{figure*}

The previous distance functions are shown in Fig.~\ref{fig:case_unitary}(a) as a function of $\epsilon$ for several $N$, where spacing factor $\ell$ has been accordingly fixed in each case according to the unitarity condition $\ell\epsilon=\frac{1}{2}\left(1-\frac{1}{N}\right)$. Singular values in $\Sigma$ are approximately close to one for half-aperture angles smaller than or around $\pi/8$, where the distance $d_{\Sigma}(\epsilon,N)$ vanishes. We can still obtain a fairly acceptable unitary approximation even for larger values in the interval $\epsilon\in(\pi/8,\pi/4)$. The latter is reinforced by computing $FF^{\dagger}$, which is equal to the identity if $F$ is unitary and $d_{F}(\epsilon,N)$ vanishes. The latter is portrayed in the inset of Fig.~\ref{fig:case_unitary}(a), where a good match with the previous measure exists, as $FF^{\dagger}$ behaves as a unitary matrix for $0<\epsilon<\pi/4$ with a high level of accuracy. Clearly, for $\epsilon>\pi/4$, both distance measures strongly deviate from zero since the assumption $\sin\theta_{n}\approx\theta_{n}$ ceases to be valid. 

Despite the accuracy of the unitary approximation, the resulting required values of $\ell$ do not lead to a one-to-one function $\Omega(\delta;\ell)$, and thus continuous tracking is ruled out for this approximation. In turn, the discrete tracking measure $\Phi(\delta,\ell)$ still reproduces the required dynamics, illustrated in Fig.~\ref{fig:case_unitary}(b)-(c) for $N=7$ and $N=20$, respectively. Here, the angle measurement is shown for $\epsilon\in(0.5,1)$ for illustration purposes. Indeed, the unitarity approximation does not hold for $\epsilon\approx 1$, but it is valid for $\epsilon$ around $0.5$. Although the detection aperture is small for $\epsilon=0.5$, the distance factor $\ell\approx 0.86$ regardless of the number of ports. Likewise, one can increase the detection aperture to $\epsilon\approx 0.625$ so that $d \approx$ 968 nm. The latter are realistic implementations for photonic circuits based on telecommunications standards composed of waveguides of 500 nm in width and an infrared light source of 1550 nm. In this case, the continuous grating coupler separation required for the architecture becomes 1333 nm and 968 nm, both larger than the waveguide width. Even though these results hold for $N=7$ and $N=20$, we have a better detection scheme for $N=20$, as the detected angle is discretized into finer samplings. Notice that, by increasing $\epsilon$, the unitary approximation loses accuracy and also leads to smaller grating coupler separation, some even below the waveguide width, a rather unrealistic scenario.
%
%---------------------------------------------------------------------
\subsection{Effects of noisy wavefronts}
So far, the proposed direction finder device's performance has been tested without considering the effect of noise. In practice, however, noise may superpose the incoming wave signal, thus, an error analysis is required to study any potential deviations. In the case of thermal noise, e.g., from solar radiation, one can consider additive noise on each port of the proposed device to perform a noise analysis. In that case, a proper analysis of the signal-to-noise ratio would be device-specific and requires knowledge of the devices' bandwidth, which is predominantly dictated by the grating couplers and waveguide arrays involved. In particular, the grating couplers act as narrow filters for the noise power density at each device port. Here, without entering the specific photonic design aspects, we provide a general noise analysis by considering random perturbations of the signal detected at each device port. Thus, we consider the perturbed input wavefront as follows
\begin{equation}
\mathbf{x}_{\gamma}(\theta) := \mathbf{x}(\theta) + \frac{\gamma}{\sqrt{M}}\mathbf{x}_{N} ,    
\end{equation}
where, $\mathbf{x}_{N}\in\mathbb{C}^{M}$ and $\gamma$ is a perturbation strength parameter, while the vector $\mathbf{x}_{N}$ has components whose real and imaginary parts are independently and identically distributed across a normal distribution $\mathcal{N}(\mu=0,\sigma=1)$.

For the sake of arbitrariness, the set of 100 random input vectors $\mathcal{S}_{\gamma}(\theta):\{\mathbf{x}^{(k)}_{\gamma}(\theta)\}_{k=1}^{100}$ is generated for each incident angle $\theta$ and for several perturbation strengths $\gamma$. We compute the mean and standard deviation of the processed discrete tracking function $\overline{\Phi}_{\gamma}(\theta;\ell):=\mu( \max{}(\widetilde{F}^{(u)}\mathbf{x}^{(k)}_{\gamma}(\theta))$ and $\Delta{\Phi}_{\gamma}(\theta;\ell):=\sigma ( \max(\widetilde{F}^{(u)}\mathbf{x}^{(k)}_{\gamma}(\theta))$, respectively. The latter is depicted in Fig.~\ref{fig:case_unitary}(d)-(e) for $\epsilon=3/4$ and equivalently $\ell=\frac{1}{2\epsilon}(1-\frac{1}{N})$ so that the unitary operator $\widetilde{F}^{(u)}$ becomes an accurate approximation of the direction-finding device. Particularly, one can see in Fig.~\ref{fig:case_unitary}(d) that the ladder-like pattern of the ideal case $\gamma=0$ is still present for $\gamma\neq 0$. Yet, minor deformation of the ladder pattern can be observed as $\gamma$ increases. This is better illustrated in Fig.~\ref{fig:case_unitary}(e), where for $0<\gamma<0.48$, the standard deviation is smaller than $0.5$. This means that any measurement of the discrete tracking function would have a potential error of no more than $\pm 1$ with respect to the ideal case $\Phi(\theta;\ell)$. Consequently, for wavefronts with an incident angle $\theta$, we would detect either the corresponding ideal discrete angle $\theta_{n}$ or one of its neighbors $\theta_{n\pm 1}$. Indeed, the angle detection error decreases for devices with a large number of ports $N$ as the discretized detected angles become finer, as illustrated in Figs.~\ref{fig:case_unitary}(b)-(c). For $\sigma\geq 0.56$, the standard deviation induces strong deviations that render detected angles beyond the vicinity of the nearest neighbor with respect to the ideal case.

%--------------------------------------------------------------------------
%%%%%%%%%%%%%%%%%%%%%%%%%%%%%%%%%%%%%%%%%%%%%%%%%%%%%%%%%%%%%%%%%%%%%%%%%%%

%%%%%%%%%%%%%%%%%%%%%%%%%%%%%%%%%%%%%%%%%%%%%%%%%%%%%%%%%%%%%%%%%%%%%%%%%%%
%--------------------------------------------------------------------------
\section{Discussion and Conclusion}
In summary, we showed that the direction-of-arrival sensing for optical waves could be formulated as a proper mapping of phase distributions onto amplitude distributions in multiport photonic devices. More importantly, such a mapping can be efficiently performed through a linear matrix operation. This was achieved by the inverse design of a programmable photonic matrix-vector multiplier device structure based on a set of input and output relations. In doing so, there is a degree of arbitrariness in defining the input/output relations. In this work, a simple set of rules was imposed to map a discrete set of plane waves with contiguous equidistant angles onto single-channel intensity peaks in the output. This simple rule straightforwardly leads to the inverse operator required for such a task in the form of a Non-uniform Discrete Fourier Transform (NUDFT), the inverse of which is ensured from the properties of Vondermonde matrices. One of the main drawbacks of this approach lies in the non-unitarity of the final transformation. Despite the latter, the device can be implemented through all-photonic components by decomposing it into its singular value decomposition. This permits the identification of three components, two unitary parts, and an amplitude modulator in the form of a diagonal matrix. Clearly, this prohibits intensity preservation and, thus, the use of active elements to scale up or attenuate the intensity depending on the case under consideration.

The main advantage of our architecture lies in the capacity to perform continuous tracking of the incident plane wave angles, a task possible by analyzing the intensity pattern at the output. This reveals a distribution-like pattern whose peak moves with the incident wave angle, suggesting the existence of a mathematical operation to track such an angle. Although there is no unique way to perform such an operation, it was found that the position average operation allows the desired tracking. For this, it is required that the separation between the arrays of grating couplers remain in the sub-wavelength regime, as for large values, the continuous measure function~\eqref{angle-cont} ceases to define a one-to-one mapping. Nevertheless, in the latter case, one can define a discrete tracking measure, i.e.~\eqref{angle-disc}. This becomes a valuable resource when the number of output ports is relatively large, for the accuracy increases with $N$, which was illustrated for $N=7$ and $N=20$.

We showed that the proposed non-unitary photonic processor can be approximated to a unitary by identifying the appropriate separation distance between the grating coupler arrays and for certain detection aperture angles. This is achieved when the non-uniform sampling $\psi(\theta_{q})$ becomes a linear function of the port number $q\in\{1,\ldots,N\}$. However, this compromise generally reduces the capabilities of the device, limiting the detection aperture angle to a maximum interval of $(-\pi/4,\pi/4)$ and making continuous tracking unfeasible beyond a certain grating coupler spacing. Nonetheless, the approximation does not restrict the total number of ports $N$, allowing for improved detection accuracy. Despite the approximation trade-off, the resulting unitary device is reliable within its range of applicability, and its construction requires less than half of the number of components compared to its non-unitary counterpart and, does not require amplitude modulators, thus significantly reducing the overall footprint of the device.

Although the focus of this work is on direction-of-arrival finding, the device concept and theoretical framework presented here can also be applied to the reverse operation, i.e., beam steering. This framework can be generalized for use with programmable photonic circuits to develop photonic integrated circuit LiDAR systems/subsystems. These systems offer greater flexibility compared to conventional designs based on phased arrays (PAs) and focal plane arrays (FPAs). It remains to explore features such as size and weight, power consumption, resolution, and field of view for such programmable photonic LiDAR systems.

\begin{backmatter}

\bmsection{Funding} 
This project is supported by the U.S. Air Force Office of Scientific Research (AFOSR) Young Investigator Program (YIP) Award\# FA9550-22-1-0189, and the U.S. National Science Foundation (NSF) FuSe grant \# 2329021.

\bmsection{Disclosures} 
The authors declare no conflicts of interest.

\bmsection{Data availability}
Data underlying the results presented in this paper are not publicly available at this time but may be obtained from the authors upon reasonable request.

\end{backmatter}

%%%%%%%%%% If using BibTeX:
\bibliography{biblio}

\end{document}